# MBDRes-U-Net: Multi-Scale Lightweight Brain Tumor Segmentation Network

Longfeng Shen [1,2,3*], Yanqi Hou [1, 2,†], Jiacong Chen [1,2], Liangjin Diao [1,2], and Yaxi Duan [1,2]

[1] Anhui Engineering Research Center for Intelligent Computing and Application on Cognitive Behavior (ICACB), College of Computer Science and Technology, Huaibei Normal University, Huaibei, Anhui, China
[2] Institute of Artificial Intelligence, Hefei Comprehensive National Science Center, Hefei, China
[3] Anhui Big-Data Research Center on University Management, Huaibei, Anhui, China

*Corresponding author(s). E-mail(s): longfengshen521@126.com
†These authors contributed equally to this work.

## Abstract

Accurate segmentation of brain tumors plays a key role in the diagnosis and treatment of brain tumor diseases. It serves as a critical technology for quantifying tumors and extracting their features. With the increasing application of deep learning methods, the computational burden has become progressively heavier. To achieve a lightweight model with good segmentation performance, this study proposes the MBDRes-U-Net model using the three-dimensional (3D) U-Net codec framework, which integrates multibranch residual blocks and fused attention into the model. The computational burden of the model is reduced by the branch strategy, which effectively uses the rich local features in multimodal images and enhances the segmentation performance of subtumor regions. Additionally, during encoding, an adaptive weighted expansion convolution layer is introduced into the multi-branch residual block, which enriches the feature expression and improves the segmentation accuracy of the model. Experiments on the Brain Tumor Segmentation (BraTS) Challenge 2018 and 2019 datasets show that the architecture could maintain a high precision of brain tumor segmentation while considerably reducing the calculation overhead.Our code is released at https://github.com/Huaibei-normal-university-cv-laboratory/mbdresunet

## 1. Introduction

The glioma, which can be categorized into low-grade glioma (LGG) and high-grade glioma (HGG) subtypes, is the most common primary malignant brain tumor. It has a high incidence, recurrence rate, and mortality, but a low cure rate, which makes treatment challenging. Treatment also requires accurate medical imaging of the tumors and the processing and analysis of images, which rely heavily on physicians. Consequently, the accurate segmentation of brain tumors is key for medical diagnosis and pathological analysis, as shown in **Fig 1(e)**. This task includes the division of several subareas—that is, the enhanced core (EnC), peritumoral edema (PTE), and non-enhanced core (NEC) areas [1]. Among the many medical imaging methods, magnetic resonance imaging (MRI) offers considerable advantages in the treatment of gliomas. It can provide extensive tumor information and is non-invasive; moreover, it does not expose the patient's body to radiation during imaging. There are four common modalities (**Fig. 1(a)–1(d)**)—that is, fluid-attenuated inversion recovery (FLAIR), T1-weighted (T1), contrast-enhanced T1-weighted (T1c), and T2-weighted (T2) modalities. Notably, different imaging effects emphasize different tissue characteristics and tumor diffusion regions [2]. Doctors usually manually mark a large number of MRI scans, layer-by-layer and piece-by-piece; the division of tumor regions is dependent on their experience and expert cognition. This work is tedious, time-consuming, and prone to differences of opinion; therefore, accurate automated segmentation has important research implications.

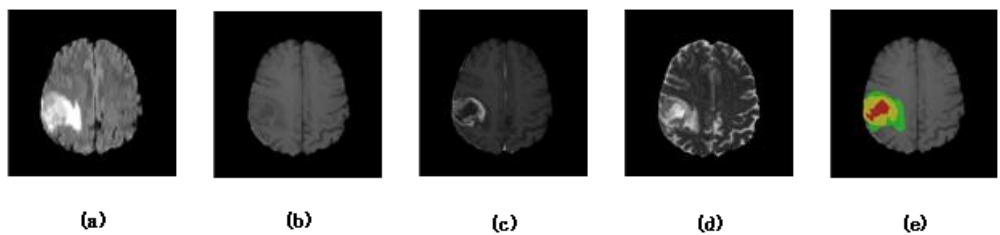

**Fig. 1** Visualization of a random sample in the Brain Tumor Segmentation (BraTS) 2018 dataset. From left to right: (a) FLAIR, (b) T1, (c) T1c, (d) T2, and (e) ground truth. Red denotes the necrotic and non-enhancing core, yellow denotes the enhanced core, and green denotes the peritumoral edema

Many attempts have been made to solve the multimodal MRI brain tumor segmentation problem, including edge segmentation, region-based [3], atlas-based [4], and machine learning-based methods. Although they have low computational complexity, these methods rely heavily on artificial annotation and a large number of datasets for pre-training, which can be time-consuming and result in poor segmentation performance.

Since 2012, the multimodal brain tumor segmentation (BraTS) challenge (jointly sponsored by the International Association for Medical Image Computing and Computer-Assisted Interventions (MICCAI) and other organization), has highlighted the progress of deep-learning algorithms in brain tumor segmentation tasks. The performance of U-Net [5] and its variants in multimodal brain tumor segmentation tasks has made considerable progress. From the three-dimensional (3D) U-Net model [6] to the TransBTS [7] and SwinUNETR models [8], studies have continuously improved the performance of brain tumor segmentation methods. Moreover, from the initial integration of the U-Net architecture to various attempts to introduce transformers, these models have introduced innovation and progress to the field of brain tumor segmentation. NN-U-Net [9] confirmed the prominent role of the U-Net architecture in image segmentation tasks, with the basic U-Net architecture being effective and competitive across multiple BraTS tasks.

Moreover, because the attention mechanism enables neural models to focus accurately on all relevant elements of the input information, it has become an important component for improving the performance of deep neural models. For example, the squeeze-and-excitation (SE) module in SENet

[10]—the winner of the ImageNet Competition in 2017—assigned different weights to different positions of the image from a channel domain perspective using a weight matrix to obtain more important channel-feature information. However, compared to the attention mechanism that focuses only on the channel, a module that combines the channel and spatial attention mechanism can achieve better results. For example, the convolutional block attention module (CBAM) [11] successively infers an attention diagram along the channel and spatial dimensions.

Although 3D convolutional neural networks (CNNs) have achieved significant results in brain tumor segmentation, these models are usually too complex to calculate and generate a large number of parameters, incurring a high computational overhead. Attempts have been made to alleviate this problem by using a lightweight model architecture; however, the segmentation performance of lightweight models falls short of that achieved by advanced models. Improving segmentation accuracy while maintaining a lower computational cost remains a difficult problem to solve.

Accordingly, we proposed a multimodal-based 3D brain tumor lightweight model (MBDRes-U-net), which not only solves the problem of single features based on single-modal image representation but also relieves the heavy computational burden caused by complex models. A residual block based on multibranch parallel convolution was used in the proposed model to replace the common 3D convolution block; the computational complexity was reduced using group convolution. An adaptive 3-D dilation convolution operation was introduced in the encoder to obtain multiscale feature representation, and a 3D multi-attention module (SCA3D) was added to the encoding stage to enable CNNs to focus attention on the tumor area. The proposed model exhibits specific advantages over other submodels of the same type in terms of parameter quantity and model complexity. We evaluated the model on the BraTS 2018 and BraTS 2019 datasets, obtaining good segmentation performance on both. The contributions of this study can be summarized as follows:

1) In this study, a 3D lightweight codec model, MBDRes-U-net, was proposed for the multi-modal MR image segmentation of brain tumors. MBDRes-U-net was developed on a symmetric encoder–decoder architecture that integrated the new residual block and attention mechanism. With its simple model structure, it can be used as a baseline for 3D brain tumor segmentation, thus promoting research on brain tumor MRI image segmentation.
2) A new residual block based on a multibranch atrous convolution was proposed, which simultaneously enlarged the acceptance domain and reduced the number of parameters, thereby solving the computational overhead problems caused by 3D convolution. Additionally, we introduced adaptive atrous convolution instead of an ordinary atrous convolution, enriching the feature representation.
3) A plug-and-play 3D attention module (3D SACA), better suited to brain tumor segmentation, was proposed. An improved combination mechanism between the spatial and channel attention was explored.

## 2. Related Work

### 2.1 Convolutional Neural Network Model

In recent years, deep learning methods, particularly CNNs, have been widely used in medical image processing applications. CNN-based models can accurately capture the local features of two-dimensional (2D) and 3D medical images through learning. Kamnitsas *et al.* proposed a fully connected multiscale model architecture (DeepMedic) [12] to segment 3D brain tumor images. Chen *et al.* proposed improving the segmentation accuracy of brain tumors using the dense connection of a CNN [13]. With the emergence of U-Net [5], the advantages of the encoder–decoder symmetric architecture (based on skip links) in medical image segmentation have become increasingly evident, and it has been widely used in brain tumor segmentation in recent years. However, the data loss caused by using 2D slices of medical images can be difficult to ignore. To better learn the imaging characteristics of 3D data and meet the clinical accuracy requirements for providing medical assistance, Çiçek *et al.* extended the U-Net model from 2D to 3D images [14]. Subsequently, Wang *et al.* proposed a brain tumor segmentation model based on the 3DU-Net model [15]. Recently, Jiang *et al.* proposed the two-stage cascaded U-Net model that won first place in the BraTS 2019 Challenge Segment Task [16]. TransBTS [7] replaces the 3DU-Net [14] bottleneck layer with a transformer integrated block to extract more full-text information and compensate for the inability of the earlier model to establish long connections; the participating team from the German Cancer Research Center proposed a brain tumor segmentation method based on the nnU-Net model, winning the BraTS 2020 Challenge segmentation task [9]. Hatamizadeh *et al.* proposed a full transformer encoder model based on the U-net architecture—namely the SwinUNETR [8] model—which has had a considerable impact on brain tumor segmentation. The winning model for 2021 was an improvement based on the nnU-Net model [17].

With the rapid development of CNN technology, the improving computational cost of models is evident. Considering the efficiency of CNN-based models, many lightweight models for brain tumor segmentation have been proposed. Chen *et al.* used a separable 3D convolution (S3D-U-Net) to reduce the computational cost and memory requirements [18]; however, the segmentation accuracy, particularly for enhancing tumor regions, was low. Zhou *et al.* [19] proposed a 3D residual neural model (ERV-NET) using a lightweight ShuffleNetV2 model [20] as the encoder and introduced a residual block decoder to avoid degradation. Chen *et al.* proposed a new 3D extended multi-fiber model (DMF-Net) in which the multi-scale image representation for segmentation was obtained by introducing a weighted 3D extended convolution operation, which reduced the number of model parameters and achieved accurate segmentation [21]. However, the lack of channel information exchange has not yet been resolved. The HMNet model [22] used high-resolution multiresolution branches in parallel to extract multiresolution features, further reducing the complexity and computational overhead of the model.

**2.2 Attention Mechanism**

To further improve the accuracy of tumor segmentation models, attention mechanisms can be introduced to focus attention on tumor-related regions. The most commonly used attention mechanisms in medical image processing include channel and spatial attention, both of which enhance the original features by aggregating the same features in all locations using different aggregation strategies, transformations, and enhancement functions. For example, Mobarakol *et al.* adopted a 3D U-Net architecture that combined channel and spatial attention with a decoder model for segmentation [23]. The MBANet model [24] included 3D multibranch attention using 3D spatial attention (SA) as the attention layer in the encoder to offer channel and spatial attention. Inspired by SA, the 3D SACA attention module used a channel shuffle [20] to promote information flow between channels without generating parameters. However, the initial grouping operation was discarded and 3D SE was used as the 3D channel excitation module and the residual block with 3D convolution as the spatial excitation module. As a plug-and-play attention block, the 3D SACA module is more suitable for processing 3D convolution models of 3D images.

## 3. Methods

### 3.1 MBDRes-U-Net

The architecture of the MBDRes-U-Net model is shown in **Fig. 2**. The algorithm uses the codec framework of the 3D U-Net architecture. Considering that the segmentation target tumor region is located in the MRI output, we added a multibranch 3D SACA mixed attention module in front of the encoder so that the model could focus more attention to the region of interest before extracting features.

During the encoding stage, a $3 \times 3 \times 3$ convolution with a step size of 2 was first used for downsampling to reduce the display memory, after which multi-scale features were extracted through six multi-branch extended convolution residual (MBDRes) blocks with adaptive extended convolution layers. In the decoder, a trilinear interpolation method was used to up-sample the feature maps. These up-sampled feature maps were then concatenated with high-resolution features obtained from skip connections. Subsequently, the concatenated features was passed through a decoding convolution block followed by an MBRes block to progressively recover the original resolution step-by-step. Next, all the channel information was fused using a $1 \times 1 \times 1$ convolution with a step size of 1. Finally, a segmentation map was obtained using the SoftMax function to realize end-to-end segmentation.

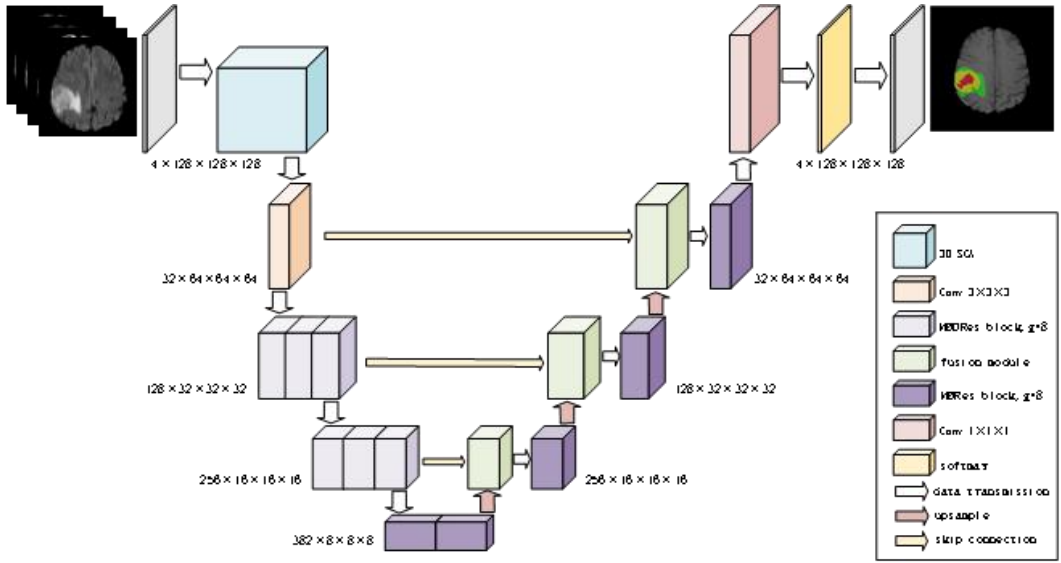

**Fig. 2** Proposed multi-scale lightweight model (MBDRes-U-Net) for MRI segmentation of brain tumors, where the size corresponds to $Channel \times Height \times Width \times Depth$, g denotes the number of branches, and g = 8 in this work

### 3.2 Multibranch Residual Block

The era of 3D CNNs enabled the full utilization of the characteristics of MRI 3D data. However, when the 3D convolution kernel runs on the entire feature mapping channel, the computational complexity—that is, the floating-point operations per second (FLOPS)—grows exponentially. Consequently, a 3D CNN incurs a high computational cost during training.

After the pre-activated residual block (**Fig, 2(a)**) used in ResNET v2 [25] is extended to a 3D convolution model (**Fig. 2(b)**), group convolution—an effective model acceleration method (including ResNeXt [26] and ShuffleNet [27] models)—can be introduced to alleviate the computational burden. Suppose that the ResNET v2 unit [25] is divided into $g$ parallel branches, and the kernel size is constant at $3 \times 3 \times 3$, then the original parameter quantity of ($b$) is $\text{Params}(b) = 3 \times 3 \times 3 \times ((C_{in} \times C_{mid} + C_{mid} \times C_{out}))$. The parameter quantity of the residual block after multi-branch grouping is $\text{Params}(b`) = g \times 3 \times 3 \times 3 \times ((C_{in}/g \times C_{mid}/g + C_{mid}/g \times C_{out}/g)) = \text{Params}(b`)/g$, which is reduced by $g$ times.

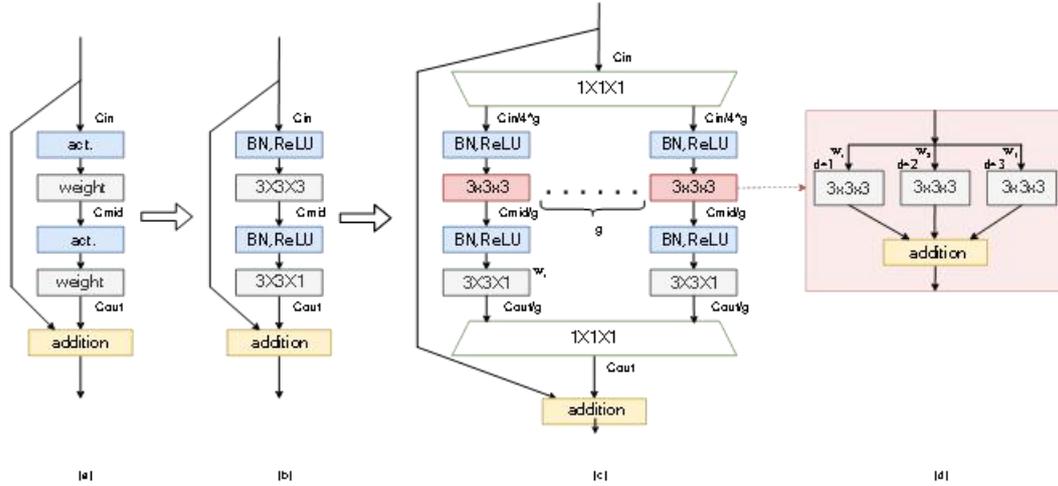

**Fig. 3** Structure of the multi-branch residual block. The parameter $g$ denotes the number of branches, $g$ = 8 in this work. The parameters $w_1$, $w_2$, and $w_3$ denote the weights of each branch of the adaptive dilation convolutional layer, and $d$ denotes the dilation rate

This grouping strategy can effectively reduce the number of model parameters and accelerate the calculations. However, the multiple branches generated by channel grouping work independently and in parallel, affecting the normal information interaction between channels and reducing the learning ability of the model. To solve the problem of a lack of channel information exchange (**Fig 2(c)**), we can add a $1 \times 1 \times 1$ convolutional layer at the beginning and end of the residual block to serve as the information route between each branch. Additionally, the residual connection can be arranged outside the unit such that information at a lower level can be transmitted directly to a higher level without generating additional parameters, increasing the learning ability of the model. Thus, the MBR block can be denoted an MBRes block.

Considering that the size of the convolution kernel in the traditional convolution is limited (which leads to its limited acceptance domain), to expand the receptive field of the model, learn the multiscale features of brain tumor MRI, and capture 3D spatial correlations, we introduced an adaptive weighted expanded convolution layer in the encoder part to replace the conventional convolution operation; thus, we obtained a multibranch expanded convolution residual block (denoted the MBDRes block). Capturing multiscale information is an effective strategy that has been used successfully before. Tokunaga *et al.* proposed a semantic segmentation task in pathology using three parallel CNNs and weighted concatenation to extract multiscale information [28]. In the dense fused maxout network (DFMN) [21] proposed by Chen *et al.*, three parallel convolution layers with different expansion rates were used as the weighted sum.

The structure of the adaptive weighted expansion convolution layer comprises three parallel 3D expansion volume integration branches, the expansion rates of each branch being 1, 2, and 3, respectively (**Fig. 3(d)**). Three weights ($w_1$, $w_2$, and $w_3$) are assigned to each branch after initialization, and the results of each branch are then added. The initialization of this weight ensures that each branch has the same impact on the model initially.

### 3.2 3D SACA

There is a considerable imbalance in the BraTS dataset, in which tumor regions account for only 1.5% of the MRI images and enhanced tumors (ETs) account for only 11% of the whole tumor (WT) images [29]. To eliminate the impact of large-area backgrounds on the segmentation, the learning ability of the model between the spatial details and advanced morphological features was balanced, the model being more focused on the tumor region.

A 3D attention mechanism can be introduced, and the 3D space and channel attention can be extracted using the feature relationship between the 3D space and channel, as follows:

$$X\_c' = X\_c \times \sigma\left(F_c(avgpool(X\_c))\right) \quad (1)$$

$$X\_s' = X\_s \times \sigma(F_c(X\_s)) \quad (2)$$

Its structure is shown in **Fig. 4**. A channel split [20] operation can be introduced to divide the input feature $X \in R^{c \times h \times w \times d}$ into two parts by channel, $X\_c \in R^{\frac{1}{2}c \times h \times w \times d}$ and $X\_s \in R^{\frac{1}{2}c \times h \times w \times d}$, where C, H, W and D denote the number, height, width, and depth of channels in the feature map, respectively. $X\_c$ first carries out average pooling to obtain the global channel information, before passing the information into the channel excitation module to obtain $\frac{1}{2}c \times 1 \times 1 \times 1$ channel correlation. $X\_s$ is introduced into the spatial excitation module, and the spatial feature correlation is aggregated to $\frac{1}{2}c \times 1 \times 1 \times 1$ dimension through a point multiplication operation of the 3D convolution layer, so that spatial attention weighting can be realized.

Consequently, the model is able to adaptively adjust the feature responses at different spatial positions.

Next, $X\_c'$ and $X\_s'$ are aggregated. We can then fuse the residuals to reduce the sparsity caused by the parallel excitation. Finally, channel shuffling is used to solve the Information exchange problem caused by the branch strategy.

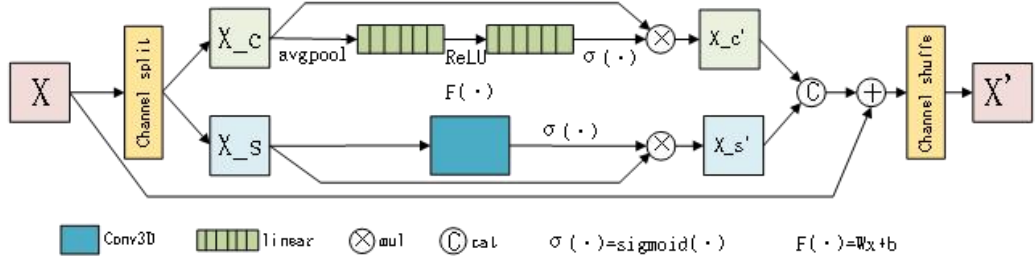

**Fig. 4** 3D SACA module structure

## 4. Results and Discussion

### 4.1 Experimental Details

#### 4.1.1 Dataset and data pretreatment

We evaluated the MBDRES-U-NET model on the following datasets:
1) BraTS 2018 dataset: The training dataset comprised 285 samples; the validation dataset comprised 66 samples.
2) BraTS 2019 dataset: The training dataset comprised 335 samples; the validation dataset comprised 125 samples.

The volume of each was $240 \times 240 \times 155$. The tags for tumor segmentation included the background (tag 0), necrotic and non-enhancing tumors (tag 1), peritumoral edema (tag 2), and GD-enhancing tumors (tag 4). To facilitate model training, the datasets were pretreated as follows:

(i) The datasets were acquired by multiple mechanisms resulting in uneven intensity; hence, we standardized the MRI images using the Z-score method.

(ii) The background information of the brain tumor image is meaningless for segmentation; therefore, we randomly cropped the data to $128 \times 128 \times 128$ voxel inputs.

(iii) To prevent over-fitting, we used the following data enhancement strategies to add training data: 0.5 for axial, coronal, and sagittal random reversals; the random rotation angle interval was $[-10°, , +10°]$.

#### 4.1.2 Assessment indicators

The effectiveness of the model was evaluated based on the computational complexity and segmentation accuracy. Validation was conducted using the validation dataset through an online portal provided by the organizers of the BraTS Challenge. Specifically, the segmentation accuracy was measured using the

Dice coefficient and Hausdorff distance (95%), where ET, WT, and TC refer to the regions of the enhanced tumor (label 1), entire tumor (labels 1, 2, and 4), and tumor core (labels 1 and 4), respectively. Complexity can be determined using the *Params* and *FLOPS* metrics. *Params* represents the spatial complexity of the model, and *FLOPS* represents its time complexity, expressed as follows:

$$\text{Params} = k_h \times k_w \times k_d \times C_{in} \times C_{out} \qquad (3)$$

$$\text{FLOPS} = 2 \times (k_h \times k_w \times k_d \times C_{in}) \times C_{out} \times h \times w \times d \qquad (4)$$

where $k_h, k_w,$ and $k_d$ denote the height, width, and depth of the convolutional kernel, respectively, $C_{in}$ and $C_{out}$ denote the number of input and output channels, respectively, and parameters *h, w,* and *d* denote the height, width, and depth of the image, respectively.

**4.1.3 Experimental setup**

We ran the experimental code in Python 3.6 using 16 lots, trained the model for 500 cycles on three parallel NVIDIA A30 GPUs, and built all experimental models using the PyTorch framework. The Adam optimizer was used, and the learning rate was set to 0.001.

**4.2 Comparison Experiments with State-of-the-art Methods**

To verify the performance of the proposed model, we compared the MBDRes-U-Net segmentation performance with that of other advanced models (including the 3D U-Net, U-Net-based, CNN with Transformer, and other lightweight models) on the BraTS 2018 and 2019 datasets. The comparison results are presented in **Tables 1** and **2**. Compared to the non-lightweight models, the proposed model exhibits the advantage of low model complexity, enabling more efficient segmentation of the WT and TC.

The parameters of MBDRes-U-Net parameters are one-fourth of those of the traditional 3D U-Net model (**Table 1**). Moreover, the computational complexity is reduced by 1643.75 G, and the segmentation accuracy is considerably improved (being 3.2%, 1.8%, and 13.6% higher than that of the 3D U-Net model in the ET, WT, and TC segmentation, respectively). Compared with those of the 3D-ESP-Net and S3D-U-Net models, although the number of parameters for MBDRes-U-Net increase marginally, the computational complexity is just one-third of both, and the segmentation accuracy is considerably improved. In comparison to the DMF-Net model with similar parameters, while the Dice coefficient of the MBDRES-U-net model is 0.1% lower, the Dice coefficients of the WT and TC increases by 0.7% and 1.8%, respectively. Although the proposed model is not as lightweight as the HMNet model, the Dice coefficients of the ET, WT, and TC increase by 0.5%, 0.2%, and 1.0%, respectively, and the Hausdorff distance reduces by 0.002, 0.389, and 2.037 mm, respectively. The overall mean score is 0.6% higher than that of the HDC-Net model, which has approximately the same computational complexity. Compared to that of the latest ADHDC-Net brain tumor segmentation model, although the score of the MBRes-U-Net model in ET is 0.3% lower, those of the WT and TC are 0.5% and 0.3% higher, respectively, the average score of the MBRes-U-Net model being 0.17% higher than that of the non-lightweight ADHDC-Net model. Consequently, the proposed method is a more efficient algorithm that can achieve comparable segmentation accuracy.

Table 1 Comparison of split performance using the BraTS 2018 dataset. (-) indicates no results reported. Best results in **bold**

| Model | FLOPS (G) | Params (M) | Dice_score (%) | | | Hausdor ff95 | | |
|---|---|---|---|---|---|---|---|---|
| | | | ET | WT | TC | ET | WT | TC |
| 3D U-Net [14] | 1669.50 | 16.21 | 75.9 | 88.5 | 71.7 | 6.040 | 17.100 | 11.620 |
| 3D-ESP-Net [30] | 76.51 | 3.36 | 73.7 | 88.3 | 81.4 | 5.302 | 5.463 | 7.853 |
| S3D-U-Net [18] | 75.20 | 3.32 | 74.9 | 89.3 | 83.1 | (-) | (-) | (-) |
| DMF-Net [21] | 27.04 | 3.88 | 79.2 | 89.6 | 83.5 | 3.385 | 4.861 | 7.743 |
| HMNet [22] | 129.40 | **0.80** | 78.6 | 90.1 | 84.3 | 2.699 | 4.727 | 7.731 |
| HDC-Net [32] | 25.53 | 2.95 | 80.0 | 89.7 | 83.5 | **2.403** | 5.615 | 6.227 |
| ADHDC-Net [31] | (-) | (-) | **80.2** | 90.0 | 84.3 | 2.411 | 4.779 | 6.125 |
| MBDRes-U-Net | **25.75** | 3.85 | 79.9 | **90.5** | **84.6** | 2.697 | **4.338** | **5.704** |

The results using the BraTS 2019 dataset are listed in **Table 2**. By retraining the model, the MBRes-U-Net model is lighter and more efficient than the 3D U-Net model. It has fewer parameters and higher segmentation accuracy than both the 3D ESP-Net and DMF-Net models. Compared to the latest brain tumor segmentation model, the MBRes-U-Net model achieves considerable improvements for ET and CT—specifically, 0.09% and 0.5% higher than the MBANet model, and 0.8% and 1.3% higher than the ADHDC-Net model, respectively. Although the MBRes-U-Net ET and WT Dice scores are 0.5% lower, the TC Dice_score is 1.6% higher than that of the TransBTS model. Compared with the HDC-Net model, which has a similar computational complexity, the TC advantages are evident (increasing by 2.6%), and the overall mean value increases by 1.03%. Compared to the HMNet model, although the proposed model parameters are 3M more than the HMNet model and the WT Dice_score is 0.4% lower, they are 1.1% and 0.5% higher for the ET and CT, respectively, with 103 G less computational complexity. Consequently, it is evident that the MBRes-U-Net model is more competitive than the other lightweight and non-lightweight models.

Table 2 Comparison of split performance using the BraTS 2019 dataset. (-) indicates no results reported. Best results in **bold**

| Model | FLOPS (G) | Params (M) | Dice_score (%) | | | Hausdor ff95 | | |
|---|---|---|---|---|---|---|---|---|
| | | | ET | WT | TC | ET | WT | TC |
| 3D U-Net [14] | 1669.50 | 16.21 | 73.7 | 89.4 | 80.7 | 6.41 | 12.32 | 10.44 |
| TransBTS [7] | 263.73 | 30.63 | **78.8** | 90.0 | 81.9 | 3.73 | 5.64 | 6.04 |
| 3D ESP-Net [30] | 76.51 | 3.36 | 66.3 | 87.1 | 78.6 | 6.84 | 7.42 | 9.74 |
| DMF-Net [21] | 27.04 | 3.88 | 77.6 | **90.0** | 81.5 | 2.99 | **4.64** | 7.98 |
| MBANet [24] | (-) | (-) | 78.2 | 89.8 | 83.0 | 3.08 | 5.88 | **5.09** |
| HMNet [22] | 129.40 | **0.80** | 77.2 | 89.9 | 83.0 | 4.01 | 5.21 | 6.57 |
| HDC-Net [32] | 25.53 | 2.95 | 77.7 | 89.6 | 80.9 | **2.95** | 8.38 | 7.68 |
| ADHDC-Net [31] | (-) | (-) | 77.5 | **90.0** | 82.2 | 4.36 | 4.78 | 6.15 |
| MBDRes-U-Net | **25.75** | 3.85 | 78.3 | 89.5 | **83.5** | 3.15 | 4.72 | 5.78 |

Additionally, several visualization results for the MBDRes-U-Net model are presented in **Fig 5**. The different colors represent different types of tumors—that is, the red area denotes necrotic and non-enhancing tumors, the yellow area denotes an enhancing tumor, and the green area denotes edema.

Moreover, from left to right, the segmentation results of the FLAIR, DMF-Net, HDC-Net, ADHDC-Net, and MBDRes-U-Net models are overlaid on the FLAIR image. As shown, the SGEResU-Net model can effectively segment the enhanced tumor, overall tumor, and core tumor regions.

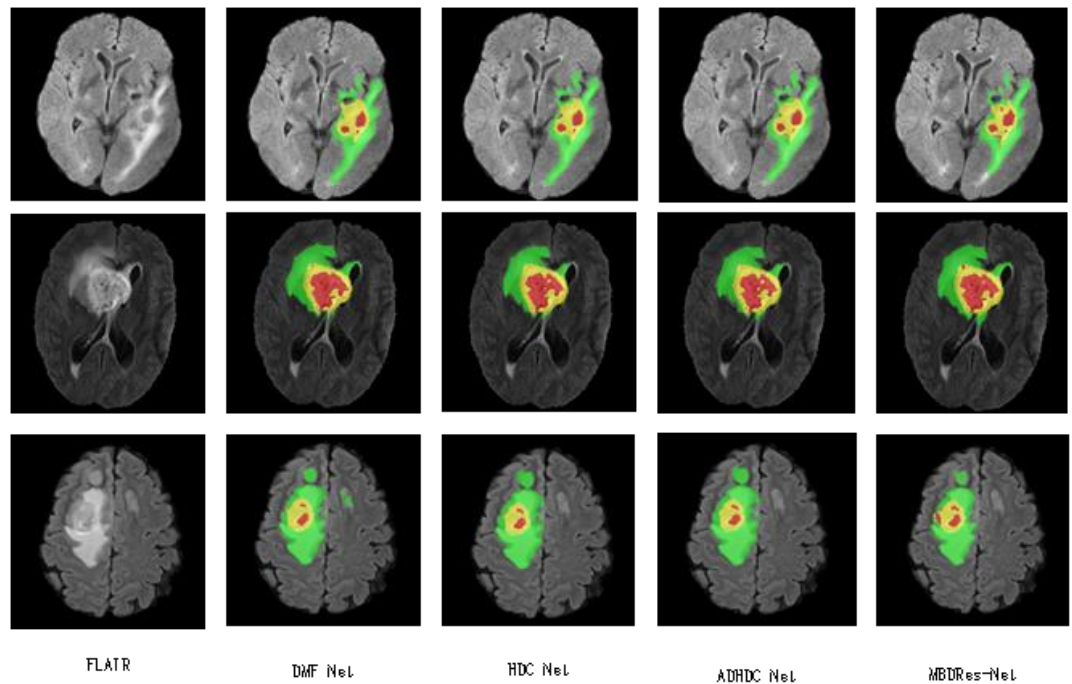

**Fig. 5** Segmentation results on the BRATS 2018 training dataset. The results are superimposed on the FLAIR image. Each color denotes a tumor category: namely red denotes necrotic and non-enhanced tumors, yellow denotes an enhance tumor, and green denotes edema

### 4.3 Ablation Experiment

#### 4.3.1 Adaptive weighted expansion convolution layer

Comparative ablation experiments, as listed in **Table 3**, were conducted to verify whether the adaptive weighted dilation convolution layer and adaptive weighting algorithm were required.

**Table 3** Comparison of different settings for the weight adaptation layer using the BraTS 2018 dataset. Best results in **bold**

| Method | Weighting | Dice_score (%) | | |
|---|---|---|---|---|
| | | ET | WT | TC |
| MBDRes-U-Net | Learnable $w_1\ w_2\ w_3$ | **79.86** | **90.45** | **84.56** |
| MBDRes-U-Net | $w_1 = w_2 = w_3 = 1$ | 78.18 | 88.96 | 83.97 |
| MBDRes-U-Net | NO Weighting | 79.31 | 89.95 | 83.98 |

Comparing the proposed model with the scheme without the adaptive weighted extended convolution layer, evidently, the extended convolution improves the Dice score. The effectiveness of the weighted strategy is proven by comparing it with an equal-weight scheme ($w_1 = w_2 = w_3 = 1$). Owing to its ability to learn and adaptively select multiscale context information, this weighting strategy results in more favorable scores, particularly for the WT metrics.

The weights $w_1$, $w_2$ and $w_3$ used in the training process are shown in **Fig. 6**. Evidently, the weight parameters in each unit are in a convergent state, reflecting the role of the multibranch extended convolution residual block. Notably, $w_1$ (blue line, corresponding to a small receptive field) plays an important role in the first three blocks and is weakened in the higher blocks. However, evidently, the extended branch of $w_3$ (red line, corresponding to the large receptive field) has a dominant influence on MBDRes Block-2–6, which may be due to the fact that the kernel with a smaller receptive field cannot capture useful semantic information at a higher level with a smaller dimension.

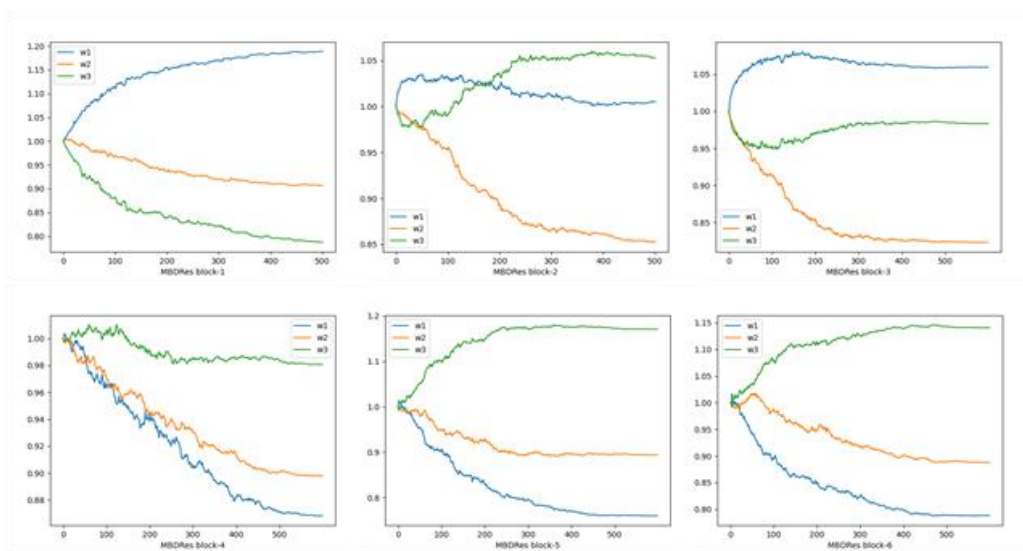

**Fig. 6** Changes of adaptive weights and sums in the training process. MBDRes block-1 is the first MBDResk block, and so on for MBDRes block-2–6. The blue line in the picture denotes $w_1$, orange denotes $w_2$ and green denotes $w_3$

### 4.3.2 Multibranch fused attention

An ablation experiment was conducted on the BraTS 2018 dataset to assess the need for attention modules and the effectiveness of the parallel branching strategy. The experimental results are summarized in **Table 4**.

**Table 4** Indicators for the evaluation of the use of different attention strategies using the BraTS 2018 dataset. Best results in **bold**

| Method | FLOPS (G) | Params (M) | Dice_score (%) | | | Hausdor ff95 | | |
|---|---|---|---|---|---|---|---|---|
| | | | ET | WT | TC | ET | WT | TC |
| No attention | | | 79.24 | 90.02 | 84.55 | 3.068 | 8.837 | 6.038 |
| ch_se | 25.745 | 3.849 | 78.24 | 52.59 | 82.55 | 2.723 | 20.013 | 6.460 |
| ch_se + sp_se | 25.753 | 3.849 | 78.78 | 89.87 | 83.31 | 2.762 | 5.376 | 6.378 |
| 3D SA | 25.741 | 3.849 | 79.12 | 89.87 | 84.32 | 2.921 | 6.147 | **5.598** |
| 3D SACA | 25.745 | 3.849 | **79.86** | **90.45** | **84.56** | **2.697** | **4.338** | 5.704 |

Using only the channel excitation module (ch_se) and the channel model in series with a spatial excitation module ( ch_se + sp_se ) verified the need for branch parallelism. Additionally, a comparative experiment was conducted using the 3D SA model instead of the 3D SACA model (which also exhibits dual-attention parallelism). The experimental results show that the proposed attention mechanism is more competitive.

**5. Conclusions**

In this study, we proposed a novel 3D U-Net model—namely, the MBDRes-U-Net model—which comprised a multibranch residual block and integrated multibranch fused attention. In particular, we placed an attention block in the encoder to ensure the model paid more attention to the region of interest before extracting the features. In the encoder, we introduced an adaptive weighted expansion convolutional layer to replace the common convolutional layer to form a multibranch expansion convolutional residual block and extract the multiscale features, enriching the feature representation of the image. Additionally, the MBDRes-U-Net model was compared with advanced methods using the BraTS dataset. Through quantitative analysis of the MBDRes-U-Net segmentation results and labels, it was evident that the MBDRes-U-Net model exhibited better computational efficiency, fewer parameters, and optimal segmentation performance. In the future, we plan to evaluate the application of the MBDRes-U-Net model to other typical medical image segmentation tasks.

**References**

[1] Menze BH, Jakab A, Bauer S, *et al.* (2014) The multimodal brain tumor image segmentation benchmark (BRATS) [J]. IEEE Trans Med Imaging, 34(10): 1993–2024.

[2] Baid, U, Ghodasara, S, Bilello, M, Mohan, S, Calabrese, E, Colak, E, Farahani, K, Kalpathy–Cramer, J, Kitamura, FC, Pati, S, *et al.* (2021) The RSNA-ASNR-MICCAI BraTS 2021 benchmark on brain tumor segmentation and radiogenomic classification. arXiv preprint arXiv:2107.02314

[3] Liu J, Li M, Wang J, *et al.* (2014) A survey of MRI-based brain tumor segmentation methods [J]. Tsinghua Science and Technology, 19(6): 578–595.

[4] Wang H, Suh JW, Das SR, *et al.* (2012) Multi-atlas segmentation with joint label fusion [J]. IEEE Trans Pattern Anal Mach Intell, 35(3): 611–623.

[5] Ronneberger O, Fischer P, Brox T (2015) U-net: Convolutional networks for biomedical image segmentation [C] // Medical Image Computing and Computer-Assisted Intervention–MICCAI 2015: 18[th] International Conference, Munich, Germany, October 5–9, 2015, Proceedings, Part III 18. Springer International Publishing, 234–241.

[6] Çiçek Ö, Abdulkadir A, Lienkamp SS, et al. (2016) 3D U-Net: Learning dense volumetric segmentation from sparse annotation [C] // Medical Image Computing and Computer-Assisted Intervention–MICCAI 2016: 19th International Conference, Athens, Greece, October 17–21, Proceedings, Part II 19. Springer International Publishing, 424–432.

[7] Wang W, Chen C, Ding M, et al. (2021) TransBTS: Multimodal brain tumor segmentation using transformer [C] // Medical Image Computing and Computer Assisted Intervention–MICCAI 2021: 24th International Conference, Strasbourg, France, September 27–October 1, Proceedings, Part I 24. Springer International Publishing, 109–119.

[8] Hatamizadeh A, Nath V, Tang Y, et al. (2020) Swin UNETR: Swin transformers for semantic segmentation of brain tumors in MRI images [C] // Brainlesion: Glioma, Multiple Sclerosis, Stroke and Traumatic Brain Injuries: 6th International Workshop, BrainLes 2020, Held in Conjunction with MICCAI 2020, Lima, Peru, October 4, Revised Selected Papers, Part II 6. Springer International Publishing, 272–284.

[9] Isensee F, Jäger PF, Full PM, et al. (2020) nnU-Net for brain tumor segmentation [C] // Brainlesion: Glioma, Multiple Sclerosis, Stroke and Traumatic Brain Injuries: 6th International Workshop, BrainLes 2020, Held in Conjunction with MICCAI 2020, Lima, Peru, October 4, Revised Selected Papers, Part II 6. Springer International Publishing,118–132.

[10] Hu J, Shen L, Sun G (2018) Squeeze-and-excitation networks [C] // Proceedings of the IEEE Conference on Computer Vision and Pattern Recognition. 7132–7141.

[11] Woo S, Park J, Lee JY, et al. (2018) CBAM: Convolutional block attention module [C] // Proceedings of the European Conference on Computer Vision (ECCV), 3–19.

[12] Kamnitsas K, Ledig C, Newcombe VFJ, et al. (2017) Efficient multi-scale 3D CNN with fully connected CRF for accurate brain lesion segmentation [J]. Med Image Anal, 36: 61–78.

[13] Chen L, Wu Y, D'Souza AM, et al. (2018) MRI tumor segmentation with densely connected 3D CNN [C] // Medical Imaging 2018: Image Processing. SPIE, 10574: 357–364.

[14] Çiçek Ö, Abdulkadir A, Lienkamp SS, et al. (2016) 3D U-Net: learning dense volumetric segmentation from sparse annotation [C] // Medical Image Computing and Computer-Assisted Intervention–MICCAI 2016: 19th International Conference, Athens, Greece, October 17–21, Proceedings, Part II 19. Springer International Publishing, 424–432.

[15] Wang F, Jiang R, Zheng L, et al. (2019) 3D U-Net based brain tumor segmentation and survival days prediction [C] // Brainlesion: 5th International Workshop, BrainLes 2019, Held in Conjunction with MICCAI 2019, Shenzhen, China, October 17, Revised Selected Papers, Part I 5. Springer International Publishing, 131–141.

[16] Jiang Z, Ding C, Liu M, et al. (2019) Two-stage cascaded U-Net: 1st place solution to BraTS challenge 2019 segmentation task [C] // Brainlesion: Glioma, Multiple Sclerosis, Stroke and Traumatic Brain Injuries: 5th International Workshop, BrainLes 2019, Held in Conjunction with MICCAI 2019, Shenzhen, China, October 17, Revised Selected Papers, Part I 5. Springer International Publishing, 231–241.

[17] Luu HM, Park SH. (2021) Extending nn-U-Net for brain tumor segmentation [C] // Brainlesion: 7th International Workshop, BrainLes 2021, Held in Conjunction with MICCAI 2021, Virtual Event, September 27, Revised Selected Papers, Part II 7. Springer International Publishing, 173–186.

[18] Chen W, Liu B, Peng S, et al. (2018) S3D-UNet: Separable 3D U-Net for brain tumor segmentation [C] // Brainlesion: Glioma, Multiple Sclerosis, Stroke and Traumatic Brain Injuries:

4th International Workshop, BrainLes 2018, Held in Conjunction with MICCAI 2018, Granada, Spain, September 16, Revised Selected Papers, Part II 4. Springer International Publishing, 358–368.

[19] Zhou X, Li X, Hu K, *et al.* (2021) ERV-Net: An efficient 3D residual neural network for brain tumor segmentation [J]. Expert Syst Appl, 170: 114566.

[20] Ma N, Zhang X, Zheng HT, *et al.* (2018) ShuffleNet v2: Practical guidelines for efficient CNN architecture design [C] // Proceedings of the European Conference on Computer Vision (ECCV), 116–131.

[21] Chen C, Liu X, Ding M, *et al.* (2019) 3D dilated multi-fiber network for real-time brain tumor segmentation in MRI [C] // Medical Image Computing and Computer Assisted Intervention–MICCAI 2019: 22nd International Conference, Shenzhen, China, October 13–17, Proceedings, Part III 22. Springer International Publishing, 184192.

[22] Zhang R, Jia S, Adamu MJ, *et al.* (2023) HMNet: Hierarchical Multi-Scale Brain Tumor Segmentation Network [J]. J Clin Med, 12(2): 538.

[23] Islam M, Vibashan VS, Jose VJM, *et al.* (2019) Brain tumor segmentation and survival prediction using 3D attention U-Net [C] // Brainlesion: Glioma, Multiple Sclerosis, Stroke and Traumatic Brain Injuries: 5th International Workshop, BrainLes 2019, Held in Conjunction with MICCAI 2019, Shenzhen, China, October 17, Revised Selected Papers, Part I 5. Springer International Publishing, 262–272.

[24] Cao Y, Zhou W, Zang M, *et al.* (2023) MBANet: A 3D convolutional neural network with multi-branch attention for brain tumor segmentation from MRI images [J]. Biomed Signal Process Control, 80: 104296.

[25] He K, Zhang X, Ren S, *et al.* (2016) Deep residual learning for image recognition [C] // Proceedings of the IEEE Conference on Computer Vision and Pattern Recognition.770–778.

[26] Xie S, Girshick R, Dollár P, *et al.* (2017) Aggregated residual transformations for deep neural networks [C] // Proceedings of the IEEE Conference on Computer Vision and Pattern Recognition, 1492–1500.

[27] Zhang X, Zhou X, Lin M, *et al.* (2018) ShuffleNet: An extremely efficient convolutional neural network for mobile devices [C] // Proceedings of the IEEE Conference on Computer Vision and Pattern Recognition, 6848–6856.

[28] Tokunaga H, Teramoto Y, Yoshizawa A, *et al.* (2019) Adaptive weighting multi-field-of-view CNN for semantic segmentation in pathology [C] // Proceedings of the IEEE/CVF Conference on Computer Vision and Pattern Recognition, 12597–12606.

[29] Wang P, Chung ACS (2022) Relax and focus on brain tumor segmentation [J]. Med Image Anal, 75: 102259.

[30] Nuechterlein N, Mehta S. (2018) 3D-ESPNet with pyramidal refinement for volumetric brain tumor image segmentation [C] // Brainlesion: Glioma, Multiple Sclerosis, Stroke and Traumatic Brain Injuries: 4th International Workshop, BrainLes 2018, Held in Conjunction with MICCAI 2018, Granada, Spain, September 16, Revised Selected Papers, Part II 4. Springer International Publishing, 245–253.

[31] Liu H, Huo G, Li Q, *et al.* (2023) Multiscale lightweight 3D segmentation algorithm with attention mechanism: Brain tumor image segmentation [J]. Expert Syst Appl, 214: 119166.

[32] Luo Z, Jia Z, Yuan Z, *et al.* (2020) HDC-Net: Hierarchical decoupled convolution network for brain tumor segmentation [J]. IEEE J Biomed Health Inform, 25(3): 737–745.